# A Theory and Calculation of Lunar Center of Mass Shift and the Crustal Thickness Difference Between Far and Near Sides of the Moon

Otto B. Bischof [1]

[1] Math Department, Laney College; obischof@peralta.edu

**Abstract:** The cause for the difference in crustal thickness between the far and near sides of the moon has been considered an open problem in astronomy since 1959 when the Soviet spacecraft Luna 3 sent back the first images of the lunar farside. The problem is referred to as the lunar farside highlands problem. In this article, the author deduces the center of mass shift of the moon from its geometrical center and the difference in crustal thickness between the far and near lunar sides necessary to explain it. Although there have been other theories proposed to explain these phenomena, this theory explains them as arising naturally from the effects on lunar material due to the earth's external gravity force and the moon's synchronous rotation and revolution. The author's mathematical model results in a calculated value of 1.6 km for the center of mass shift and a crustal thickness difference of 16 km. The author's deduced values of center of mass shift and crustal thickness difference are close to within about 5 to 10 % of their accepted values. One unexpected and deducible result of the author's theory is that the center of mass shift of the moon is decreasing towards moon center under lunar recession at the current rate of about $1.6 \times 10^{-4}$ mm/yr.

## 1. Introduction

Scientists have proposed several theories on the formation of the moon. In 1946, Reginald A. Daly became the first to suggest that the moon had formed in a collision with earth [1]. But Daly's theory didn't get the attention it deserved until later in the 1970s when William K. Hartmann and Donald R. Davis suggested that an earth-proto-planet collision early in the history of the solar system could have sent an ejection of non-volatile dust into orbit that could then re-condense to form our moon [2]. Alistair G.W. Cameron and William R. Ward contributed as well to what by then had become the leading theory of moon origin: the Giant Impact Hypothesis. They proposed specifically that a proto-planet collided tangentially with the earth, vaporizing both non-volatile, lighter materials and volatile materials, while the heavier metallic material sank into the earth. Only the non-volatile material could re-condense to form the moon; the volatiles would escape. This model explained the lunar abundance in light, non-volatile materials, its deficiency in iron, as well as its lack of volatiles relative to earth [3].

Since 1959, when spacecraft were able to orbit the moon, scientists have been mapping the moon's surface topography. Lunar scientists have combined topography measurements with measurements of lunar gravity to make models of lunar crustal thickness and deduce the values of macroscopic quantities. Smith, D.E. *et al* have measured lunar topographic features from lunar orbiters like the Lunar Reconnaissance Orbiter (LRO) using laser altimeters like LOLA [4]. Others have developed empirical crustal models to deduce macroscopic quantities like the bulk lunar crustal density from gravity and topography measurements [5]. The researchers assumed that the lunar gravity field depended on surface topography, surface basalt flows, and the crustal/mantle interface boundary [6]. After subtracting off the influence of the irregularities in topography and basalt flows, the researchers were able to uncover variations of the crustal/mantle interface, the signature of differences in crustal thickness [7]. Huang, Q. and Wieczorek, M.A. also discovered that the crustal rock density and the crustal bulk density are different because moon crust is porous [5]. To prove their assertion, they measured the densities of Apollo lunar crustal rock samples and compared them to the bulk densities of crustal material as inferred from spacecraft gravity measurements [5]. The work of these experimentalists and others like them has provided much needed knowledge of the moon. However, the knowledge has been mostly empirical and has not explained from first principles of physics alone how the moon formed.

An asymmetry in the lunar far and near sides has remained a mystery to lunar scientists since 1959, for example. Although experimentalists discovered it, they have not explained how it could have happened. Maria are greyish colored rock with greater density that may have seeped out from within the moon after collisions with asteroids early in the history of the solar system [8]. After bombardment, radioactive elements like thorium heated the inner lunar material and the pressurized material expanded into the impact basins through points in the crust that had been weakened by the impacts [8]. These former floods of inner lunar material are what we today identify as the maria [8]. The asymmetry in the far and near sides of the moon is this; more maria lie on the near side than on the far side and the far side highland rocks are less dense than mare rocks [7]. In addition, the far side of the moon is on average 15 km thicker than its near side [9]. Although lunar scientists believe that the reason there are more maria on the near side is because the near side has thinner crust [8], they have been mystified as to the cause for the thinner crust on the near side.

Some scientists have proposed theories to explain the crustal thickness difference from first principles of physics. Erik. I. Asphaug and Martin Jutzi have proposed that a previous collision with a second moon deposited an extra layer of crustal thickness on the far side [10]; Arpita Roy *et al* have proposed that "earthshine" kept the near side hotter during lunar formation, allowing more far side crust to crystallize [11]; and Garrick-Bethel *et al* have proposed that the far side crustal thickness may have been caused by "spatial variations in tidal heating" [12]. But the problem, until now, has remained an open one. There is no universally accepted theory as to why there is a crustal thickness difference between the far and near sides of the moon.

There has also been work that has linked the center of mass shift of the moon to the crustal thickness difference between the two sides. As early as 1980, E.L. Haines and A. E. Metzger empirically determined lunar crustal thickness variations and showed that they were consistent with the observed lunar center of mass shift from center [13]. Today, most lunar scientists seem to have accepted the idea that the center of mass shift of the moon from its center is related to the lunar crustal thickness difference [14]. What is lacking is a theory from first physics principles that relates these two quantities and deduces both of them.

Boris P. Kondratyev is a theorist who has investigated the center of mass shift of the moon. He proposes a theory for the eastward shift in center of mass from the earth-moon line, but not for the shift in center of mass *in* the earth-moon line [15]. He speculates that differences in tidal forces rigidly locked the lunar center of mass into its current position early in the moon's history [16]. Up to now, however, there is no calculation that shows how tides have rigidly locked the lunar center of mass into its current position and it has not even been known whether or not tides can account for the center of mass shift in the earth-moon line.

The author's new study is of interest because it is consistent with facts already known about physics and the moon and it is consistent with the Giant Impact Hypothesis. His theory invokes no ad-hoc hypotheses to explain the center of mass shift and crustal thickness difference. His theory is different than all the other models listed above that attempt to explain the asymmetry between the near and far sides of the moon. Instead of trying to explain the center of mass shift as a consequence of the crustal thickness difference, he explains the crustal thickness difference and center of mass shift as arising from the same consequence: Archimedes' Law, a law that is well established. His theory is consistent with the Giant Impact Hypothesis because both theories assume a molten moon was present in earth orbit at some time in the past. In addition, it is the first theory based on first principles of physics that the author is aware of that deduces a quantitative value for both the lunar center of mass shift in the earth-moon line and the lunar crustal thickness difference between the far and near sides. The author's theory is also of interest because it predicts an entirely new and interesting result: that the lunar center of mass shift is decreasing at a calculable rate towards lunar center under lunar recession from earth.

**2. Discussion**

*2.1. Statement of Assumptions of the Model*

1. The Moon has an approximately circular orbit around the earth and is in synchronous rotation around it. These effects are consistent with the fact that the earth has caused tidal forces on the moon.

2. Currently the moon has an approximate spherical shape. The variation in radius depends mostly on latitude, with a bulging at the equator and a contraction at the poles. The average radius of the moon is about 1737 km [4].

3. The Moon has several layers. For simplicity of calculation, the author assumes a 2 layer lunar interior consisting of a lower density crust and a lunar core. Although the lunar core may actually be made up of several parts with somewhat different densities, one may assume that it is only one part with an average density that one can calculate.

4. The author assumes that the inner core is approximately spherical in shape just like the overall moon. He reasons that the outside crust is so thin relative to the rest of the moon beneath it that it will not prevent the interior core from assuming a sphere. That is, the outside crust is too thin to deform the inner core shape from assuming what it would if there were no crust at all: a sphere.

*2.2. Summary of Calculation*

There are two parts to the calculation. First, the author calculates the shift of the center of mass of the moon from its geometrical center. Second, he calculates the difference in crustal thickness between the far and near sides of the moon, necessary to explain this shift. In the first part, the author assumes that the center of mass of the moon is in circular orbit around the earth's center. This is a reasonable approximation because the mass of the earth is approximately 81.301 times greater than that of the moon (as derived from [8] (pp. A-7 & A-8) and thus the center of mass of the earth-moon system lies *within* the earth. In fact, using the definition of center of mass, one may deduce that it lies approximately 4671 km away from earth center on average. This produces an approximate 1.2 % error, but makes the math simpler to work out. In addition, observations indicate that the deviation from a circular orbit relative to the distance from earth center is small for the moon, making a mean eccentricity of about .05 [8] (p. A-8), so that its orbit can be approximated well by a circle, a shape having eccentricity 0. Next, the author assumes synchronous rotation for the moon in its orbit around earth center. The author writes down an expression for the effective external gravitational field due to the earth at any point on the moon, including the effects of revolution of the Moon around earth and the gravity of earth, using a coordinate system at the earth-moon center of mass, assumed at earth center. Then he takes a Taylor expansion of the effective external gravity field expression around the moon's center of mass. He can assume that over the moon diameter, the distance to the earth center is approximately a constant called $R_{EM}$ and he introduces moon coordinate x measured from the center of mass of moon along the earth-moon line to compute small changes in this distance $R_{EM}$. The author assumes that over the moon diameter, the effective external gravitational field only varies as a function of the displacement $x$ from the center of mass of the moon.

One may then find a general formula for the effective external gravitational field that holds at any position on the moon that has a projected displacement $x$ along the earth-moon line from the center of mass of the moon. One keeps two terms in the Taylor expansion and notes that the expression is different on the near and far sides of the moon. One notes that this gravitational field must be zero at all times in the moon's co-moving frame when evaluated at the center of mass itself. The "centrifugal force" balances the net gravity force of the earth on the moon at its center of mass. If this were not so, then the moon would not be moving in a circle around the earth-moon center of mass.

Then the author notes that if the moon's center of mass were at its geometrical center, then to first order, the Taylor terms would cancel on the near and far sides and this would be consistent with the net effective gravity field due to the earth being zero at the lunar center of mass. But when the second order terms are also included, one sees that it will now be impossible for this net gravity field to be zero if the center of mass were at the lunar center, because the contributions to the second order terms always have a negative sign at every point on the moon. This must then mean that the center of mass of the moon must shift from its center some distance so that the net first order term will now become positive and balance the net second order term. If this happens, then the net gravity field due to earth and experienced by moon can be zero at the lunar center of mass.

The first part of the calculation then ends up with the author finding the center of mass shift of the moon from its center. To do it, the author assumes an average constant density sphere with the negative second order term acting at each point on the moon and he finds the net second order term by averaging over the lunar volume. The author then asks, by what distance would the lunar center of mass need to be displaced from its center along the earth-moon line towards earth to make the desired net first order term balance the net second order term so that the net effective gravity field acting at the lunar center of mass due to earth and experienced by the moon is zero. Here one may assume that the net second order term will change very little after the shift since one expects the shift to be small relative to the moon's radius.

The second calculation asks what is the difference in crustal thickness between the far and near sides that can be deduced from such a center of mass shift. One may assume a two-layer density structure with the light crust on the outside and an inner metallic-like core with average density greater than the crust. One knows from lunar observations that the outer moon volume is approximately spherical. One may also assume that because the mean lunar crust thickness and the differences in it between the far and near sides are so small relative to the lunar radius, that the inner core material will deform very little from the spherical shape it would take if there were no crust. A simple geometrical argument following the definition of the center of mass then allows the author to find an approximate solution to the difference in crustal thickness between the far and near sides of the moon.

Both of the calculations yield results that are very close to the lunar observational data. There is some error. However, for an astrophysics problem, in which many complicating factors are often present at once, and where theory may sometimes even be considered acceptable when it predicts a quantity to within a factor of 10, this theory and its assumptions seem likely to be true because of their close agreement with observation. As far as the author is aware, the theoretical calculations described here have never been done before and the physical and math model appears nowhere else in the literature.

*2.3. Qualitative Analysis*

The author's immediate hypothesis before performing the calculations was that the difference in crustal thickness between the far and near sides of the moon is due to a multi-density moon that at one time was molten in the effective gravity field of earth. One may think that there must have been a net effective external gravity field across the moon due to the earth. If there had been, then by the Buoyancy Law of Archimedes, the higher density material would have moved in the direction of this net external gravity field. The lower density material would have floated on "top" of it. Here, "top" means in the direction away from the direction that this gravity field points in. This buoyancy law of Archimedes is induced to explain planetary differentiation in the earth and other planets. Planetary differentiation is the process by which planets achieve a series of concentric layers, with the ones closest to the inside having the highest density. According to this theory, when planets form, the higher density material sinks in the direction of the gravity field, which points towards the center, while the less dense material floats on top. This model should be similar for the moon only the gravity field that the lunar density layers would respond to would not be due to themselves alone, but to the earth as well and the effects of revolution around it.

One may look for a net external gravity field across the moon to support this hypothesis. One may assume that the lunar center of mass would be located at its center to begin with. After taking a Taylor approximation and considering the second order approximation, one then finds that the second order term would produce a net negative gravity field across the moon in the direction towards earth because the first order term would cancel out if the moon's center of mass were coincident with its geometrical center. This supports the hypothesis. One may think that it might be the Buoyancy Law of Archimedes that explained the difference in crustal thickness.

The author then calculates the shift in center of mass of the moon from its center. If one assumes a uniform density sphere for the moon to begin with, one may find the shift in center of mass that would induce a great enough net positive contribution in the first order term to balance the net second order term. Then once one calculates the shift in center of mass, one may deduce the crustal thickness difference by making some reasonable approximations.

The author believes that the moon was once in a molten state in which the external gravity of the earth caused its higher density lunar core to move towards earth preferentially more on the near side than the far side. The lunar center of mass would then shift from its center towards earth. It would shift an amount so that the net first order term would cancel the net second order term and the effective external gravity field of earth experienced by moon at its center of mass would now be zero. This was so the moon could attain a circular orbit around earth, a trajectory that earth tidal forces would enforce. This would then "switch off" the separation of the two density layers in that configuration, because now the overall net gravity field across moon would be zero. Cooling would then occur and this would make the configuration more permanent. As the moon receded from earth, tidal forces would have decreased in magnitude and the moon would have relaxed into the more spherical shape we see today.

Moreover, the math for the author's theory gives an indication of the evolution of the lunar center of mass shift as the moon recedes from earth. The math indicates that the center of mass shift is inversely proportional to $R_{EM}$, the earth-moon distance (please see Equation 28). As $R_{EM}$ increases, the net second-order term averaged over the whole moon will become less negative (or more positive), since it depends inversely on the negative of the fourth power of the earth-moon distance (please see Equation 22). On recession, then, the earth induces an unbalanced, net second-order effective external gravity field across the moon pointing away from the earth. The math shows that the net first-order and second-order terms can only balance to make the overall net external effective gravity field zero if the center of mass shifts towards the center as the earth-moon distance increases. The physical mechanism consistent with this math model may involve the settling or relaxation of a semi-elastic or liquid lunar material in the presence of this induced effective external gravity field that points away from earth on lunar recession. Watters *et al* present evidence of 28 moonquakes measuring from between 2 and 5 on the Richter scale from seismometers left at four Apollo landing sites, confirming that this may be happening [17]. Although scientists do not know the reason for moonquakes, they believe core contraction and crust wrinkling causes them as the liquid core cools [17]. These observations are also consistent with the idea presented here of the lunar center of mass shift moving towards the lunar center. If the author's theory is correct, then these effects should occur when lunar material rearranges itself as the moon recedes from earth.

The author's theory can be tested. It predicts through equation (28), developed in the following pages that the lunar center of mass will shift towards the center of the moon as the moon recedes from earth. In fact, using the current rate of lunar recession of 3.8 cm/yr, [8] (p.160) and using the known values of lunar radius and earth-moon distance, one can differentiate equation (28), to calculate that the current rate of lunar center of mass shift towards center is $1.6 \cdot 10^{-4}$ mm/yr. Please see equations (32) and (33). If this prediction is not found to be true, then either this theory or its assumptions are false. On the other hand, if this prediction is found to be true, then any other theory that predicts a lunar center of mass rigidly locked into a fixed position long ago must be discarded.

Equation (43) indicates that the crustal thickness difference between the far and near sides of the moon is directly proportional to the center of mass shift. Thus, the author's theory also predicts that as the moon recedes from the earth over time due to tidal effects, this crustal thickness difference will decrease as the center of mass shift does.

The author finds that tides in themselves can't explain the lunar center of mass shift. Today, for example, the observed shift is about 1.8 km from the lunar center [9] (p. 229). Observed tidal bulges on the near and far sides are about 50 cm [18], however, and only the small differences in the tidal bulges on the far and near sides could explain a center of mass shift if tides were to explain the shift. The theory the author proposes in this paper is more likely then since it predicts a 1.6 km center of mass shift. It also proposes a reason why higher density material will have built up preferentially on the near side of the moon, whereas tidal theory does not.

The author could find his physical model nowhere else in the literature. As far as he can tell, this is the first time that the law of planetary differentiation and Archimedes' law have been used to predict a qualitative and quantitative result for the center of mass shift, the crustal thickness difference between the far and near sides of the moon, and the rate of change with time of the lunar center of mass shift towards moon center.

### 3. Quantitative Analysis

*3.1. Calculation of $g_{eff\text{-}ext}$, the Effective External Gravity Field at Any Position on the Moon*

Let $R_{EM}$ = the distance between the center of mass of Moon and earth.

Equilibrium in circular orbit

If the earth-moon center of mass is chosen as origin, the gravity force of earth on moon acts at the lunar center of mass as the only external force and provides the centripetal force necessary for the moon to achieve a circular orbit around earth. Both forces are proportional to the lunar mass, so the lunar mass cancels out of the equation. If the net force is evaluated in the lunar co-moving frame of reference, then the net effective external force acting at the lunar center of mass is zero, since the gravity force balances the centrifugal force. This then implies that the net effective external gravity field acting at the lunar center of mass is zero as well, since the net force is proportional to the net acceleration, $g_{net\text{-}eff\text{-}ext}$.

$$\frac{GM_E}{R_{EM}^2} = \omega_m^2 R_{EM} \tag{1}$$

$$g_{net\text{-}eff\text{-}ext} = 0 \tag{2}$$

Let $\omega_m$ = the orbital revolution and rotation speed. The two are equal, since the Moon is in synchronous rotation. Solving for $\omega_m$:

$$\omega_m = \left(\frac{GM_E}{R_{EM}^3}\right)^{\frac{1}{2}} \tag{3}$$

If R = the distance of any point on Moon from earth center and assumed earth-moon center of mass:

$$g_{eff-ext}(R) = \frac{-GM_E}{R^2} + \omega_m^2 R \tag{4}$$

This gives the effective external field on the moon as experienced at R.

Assume that the angle subtended by the Moon when observed from earth is small, so that R for any point on the Moon varies only with the displacement along the earth-moon line, and ignore any components of displacement perpendicular to it.

Equation (4) then is an expression for the effective external gravity field experienced at any point on the moon, due to the earth's gravity field and the Moon's simulated gravity field generated by accelerated revolution around the earth-moon center of mass, considered as origin.

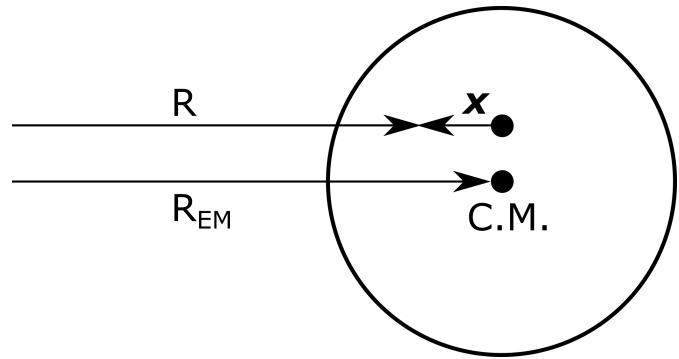

**Figure 1.** Shows vector *x* in relation to center of mass of moon.

In Figure 1 above:

Let $x$ = the displacement of any point relative to Moon's center of mass parallel to the earth-moon line. Note that if the point is on the near side of the moon's center of mass, as it is in Figure 1, then $x = -x$, while if the point lies on the far side, then $x = x$.

Next, the author finds $g_{eff}(R)$ by making a Taylor expansion around R = R$_{EM}$, the center of mass distance between earth and Moon. The author suppresses the "ext" subscript for brevity in what follows.

$$g_{eff}(R) = g_{eff}(R_{EM}) + g'_{eff}(R_{EM})x + \frac{1}{2}g''_{eff}(R_{EM})x^2 + \ldots \tag{5}$$

One may now calculate the magnitude and direction of $g_{eff}(R)$ on the near and far sides of the Moon. If R = R$_{EM}$ is substituted into equation (4) above, then:

$$g_{eff}(R_{EM}) = \frac{-GM_E}{R_{EM}^2} + \omega_m^2 R_{EM} \tag{6}$$

For a circular orbit, $\frac{GM_E}{R_{EM}^2} = \omega_m^2 R_{EM}$ \quad (1)

This implies by equation (6), that $g_{eff}(R_{EM}) = 0$.

Differentiating equation (4), and substituting in R = R$_{EM}$, one gets:

$$g'_{eff}(R_{EM}) = -GM_E(-2R_{EM}^{-3}) + \omega_m^2 \tag{7}$$

Solving equation (1) for $\omega_m^2$ and substituting the result into equation (7), one gets:

$$g'_{eff}(R_{EM}) = -GM_E(-2R_{EM}^{-3}) + \frac{GM_E}{R_{EM}^3} \tag{8}$$

Simplifying the expression on the right side of the above equation yields:

$$g'_{eff}(R_{EM}) = \frac{3GM_E}{R_{EM}^3} \tag{9}$$

If one differentiates equation (4) twice with respect to R, holding $\omega_m$ constant, and substitutes in R = R$_{EM}$, then one gets:

$$g''_{eff}(R_{EM}) = -\frac{6GM_E}{R_{EM}^4} \tag{10}$$

When these results are plugged into equation (5), one gets the effective gravity field on the near side of the Moon where $x = -x$:

$$g_{eff}(x) = 0 + \frac{3GM_E}{R_{EM}^3}(-x) + \frac{1}{2}\left(\frac{-6GM_E}{R_{EM}^4}\right)(-x)^2 + \ldots \tag{11}$$

Here, the terms decrease in magnitude as $\frac{x}{R_{EM}} \ll 1$.

Simplifying and keeping only two terms, one arrives at the following approximate formula for $g_{eff-ext}$ on the near side of the Moon where $x = -x$:

$$g_{eff-ext}(x) \approx -\frac{3GM_E}{R_{EM}^3}x - \frac{3GM_E}{R_{EM}^4}x^2 \tag{12}$$

On the far side, one has $x = x$. Plugging this and the evaluated results for the derivatives above into (5), one gets:

$$g_{eff}(x) = 0 + \frac{3GM_E}{R_{EM}^3}(x) + \frac{1}{2}\left(\frac{-6GM_E}{R_{EM}^4}\right)(x)^2 + \ldots \tag{13}$$

Simplifying the expression on the right hand side of this equation and keeping only two terms, yields the following approximate formula for $g_{eff-ext}$ on the far side of the Moon where $x = x$:

$$g_{eff-ext}(x) \approx \frac{3GM_E}{R_{EM}^3}x - \frac{3GM_E}{R_{EM}^4}x^2 \qquad (14)$$

These calculations for $g_{eff-ext}$ on the near and far lunar sides represent the effective external gravity field. The lunar gravity force is internal to the system, and its effects are not included here.

*3.2. Calculation of the Center of Mass Shift of the Moon from its Center:*

Comparison of the two fields shows a stronger magnitude gravity field on the near side than the far side. Also, one sees that if a uniform density sphere is assumed for the shape of the Moon, with a center of mass at its center, then the net first-order terms will cancel over the sphere. Since the second-order terms are always negative and point towards the earth, there is a net negative external "towards earth" gravity field across the Moon. This net negative second-order term is the reason why the center of mass must shift from the center of the Moon towards earth, as the author will show shortly.

The Moon's own gravity force on itself is a force internal to the lunar system. To any internal force acting on any Moon particle there is an equal and opposite force acting on another. This is in accordance with Newton's Third Law. Thus the sum of all the internal forces is always zero. If only internal forces act, an isolated uniform density spherical Moon will have its center of mass at its geometric center. In this configuration all the internal forces can add up to zero on all the Moon particles at once and it stays balanced.

A net force will always act at the center of mass. If the net force is internal, then it will be zero and the center of mass will not shift. The center of mass will stay at the center. The net first-order term of the external force will also act at the center of mass. The first-order term is defined:

$$\frac{3GM_E}{R_{EM}^3}x, \quad \text{if } x = x$$

$$\text{and}$$

$$\frac{-3GM_E}{R_{EM}^3}x, \quad \text{if } x = -x$$

One sees that the net first-order term is zero for a uniform density sphere, because for every positive contribution at x there will be an equal negative contribution at –x on the Moon. Thus, the net first-order term will be zero and will produce no center of mass shift from the center to first order.

However, to second-order, there will be a shift. The second order term is defined:

$$\frac{-3GM_E}{R_{EM}^4}x^2, \quad \text{if } x = x$$

$$\text{and}$$

$$\frac{-3GM_E}{R_{EM}^4}x^2, \quad \text{if } x = -x$$

So one sees that the second-order term is always negative everywhere on the Moon. It follows then that the second-order term will produce a net effective external gravity field that will shift the center of mass of the Moon from its geometrical center.

The author will now calculate the net effective external second-order gravity field acting at the center of the Moon. This is the external force that will shift the center of mass of the Moon from its center. Letting $a$ be the Moon radius and V its volume,

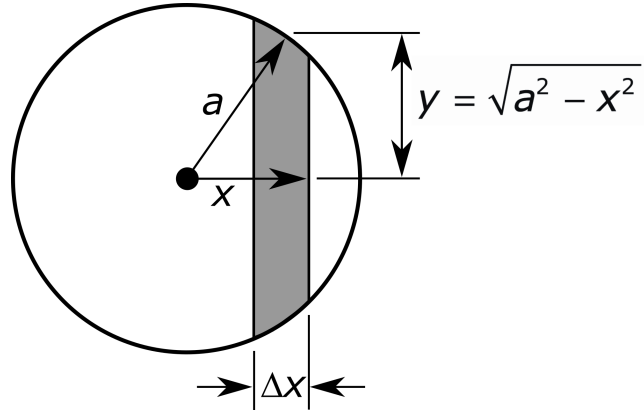

**Figure 2:** Disc volume element of the spherical moon for integration.

$$g_{eff-net-2} = \frac{1}{V} \int_{-a}^{a} g_{eff-2} dV \quad (15)$$

According to Figure 2 above, one sees that a disc element has a volume:

$$dV = \pi y^2 dx \quad (16)$$

Substitution for y as defined in Figure 2, yields:

$$dV = \pi \left(\sqrt{a^2 - x^2}\right)^2 dx \quad (17)$$

$$= \pi (a^2 - x^2) dx \quad (18)$$

Then, substituting and simplifying yields:

$$g_{eff-net-2} = \frac{1}{\frac{4}{3}\pi a^3} \int_{-a}^{a} \left(\frac{-3GM_E x^2}{R_{EM}^4}\right) \pi (a^2 - x^2) dx \quad (19)$$

$$= \frac{3}{4\pi a^3} \left(\frac{-3GM_E}{R_{EM}^4}\right) \pi \int_{-a}^{a} (a^2 x^2 - x^4) dx \quad (20)$$

$$= \frac{-9GM_E}{4a^3 R_{EM}^4} \left[\frac{2a^5}{3} - \frac{2a^5}{5}\right] \quad (21)$$

$$= \frac{-3GM_E a^2}{5R_{EM}^4} \quad (22)$$

Now, for a circular orbit:

$$g_{eff-net} = g_{eff-net-1} + g_{eff-net-2} = 0 \quad (23)$$

One sees that equations 22 and 23 imply:

$$g_{eff-net-1} = \frac{3GM_E a^2}{5R_{EM}^4} \quad (24)$$

Then one can now ask what shift $\Delta x$ in the center of mass of Moon from its center will cause this net first-order gravity field.

The first order-term $g_{eff-1}$ is defined for any point on Moon as:

$$\frac{3GM_E x}{R_{EM}^3} \quad \text{if } x = x$$

and

$$\frac{-3GM_E x}{R_{EM}^3} \quad \text{if } x = -x$$

If the center of mass of Moon were at its center, then the net first-order term would be zero. If the center of mass shifts from the center by some small amount $\Delta x$ towards earth, then the change in the first-order term $\Delta g_{eff-1}$ is for any point on Moon:

$$\frac{3GM_E \Delta x}{R_{EM}^3} \quad \text{if } x = x, \text{ since } \Delta x = \Delta x \text{ on the far side}$$

and

$$\frac{3GM_E \Delta x}{R_{EM}^3} \quad \text{if } x = -x, \text{ since } \Delta x = -\Delta x \text{ on the near side}$$

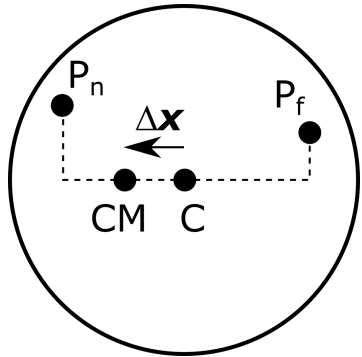

**Figure 3.** Shows center of mass shift relative to geometric center. Dotted lines indicate projected distances along the earth-moon line on near and far sides.

In Figure 3, consider a point on the near side, $P_n$. Its distance from center of mass changes due to the shift to the left by $-\Delta x$. Consider a point, $P_f$ on the far side. Its distance from the center of mass changes due to the shift by $\Delta x$.

According to Figure 3 above, one sees that the first order shift has the same positive sign everywhere on the moon.

Then, the net change $\Delta g_{eff-net-1}$ is the volume average of this term:

$$\Delta g_{eff-net-1} = \frac{1}{V} \int \frac{3GM_E \Delta x}{R_{EM}^3} dV \tag{25}$$

Since the integrand is a constant over the volume, it can come out of the integral and the integral simplifies to:

$$\Delta g_{eff-net-1} = \frac{3GM_E \Delta x}{R_{EM}^3} \tag{26}$$

One can now equate this expression to the first expression, given by equation (24) above, to deduce the center of mass shift from the center of Moon:

$$\frac{3GM_E \Delta x}{R_{EM}^3} = \frac{3GM_E a^2}{5R_{EM}^4} \tag{27}$$

This implies:

$$\Delta x_{CM} = \frac{a^2}{5R_{EM}} \tag{28}$$

When numerical values [4], [9] (pp. 228-229), are plugged into the equation, one gets:

$$\Delta x_{CM} = \frac{(1737 \text{ km})(1737 \text{km})}{5(3.844 \cdot 10^5 \text{ km})} \tag{29}$$

$$\Delta x_{CM} \approx 1.57 \text{ km} \tag{30}$$

Rounding to two significant figures, then:

$$\Delta x_{CM} \approx 1.6 \text{ km} \tag{31}$$

The quoted value for $\Delta x_{CM}$ is 1.8 km. The error in the author's result is then about 11%, when two digits are used in the calculated value of the center of mass shift. The author could find this calculation nowhere else in the literature. It is the first time, as far as he can tell, that someone has calculated a quantitative value for the center of mass shift based on a simple physical and mathematical model with the same ideas as the author.

One prediction of the author's theory is that the lunar center of mass will shift towards moon center over time. This is because the moon is now receding from earth. Differentiating equation (28) with respect to time, and plugging in known values for the lunar radius [4], the earth moon distance [9], and the current rate of lunar recession from the earth, $3.8 \frac{cm}{yr}$ [8], one arrives at the current rate of change of center of mass shift with time towards lunar center:

$$\frac{d\Delta x_{CM}}{dt} = -\frac{a^2}{5R_{EM}^2} \frac{dR_{EM}}{dt} \tag{32}$$

$$\frac{d\Delta x_{CM}}{dt} = -\frac{(1737 \text{ km})^2}{5(3.844 \cdot 10^5 \text{ km})^2}\left(38\frac{mm}{yr}\right) \approx 1.6 \cdot 10^{-4} \frac{mm}{yr} \tag{33}$$

(to two decimal places)

This prediction appears nowhere else in the literature as far as the author is aware.

*3.3. Calculation of Difference in Crustal Thickness Between Lunar Far and Near Sides:*

For this calculation, one may use known lunar quantities to deduce the crustal thickness difference given the location of the center of mass of the Moon relative to its center. One may assume that the Moon as a whole and the inner core are approximated by spheres and that an increase in the displacement of the center of the inner core towards earth relative to the Moon center will induce an increase in the observed center of mass shift that one has calculated before. One may assume further that the average density of the core, the crustal density, the overall average lunar density, the inner core volume, the total crustal volume, and the overall lunar volumes are known. This makes the overall lunar mass, the core mass, and the crustal mass all known as well. The question one may now ask is, "How far must the center of the inner core be moved from the center along the earth-moon line to make the calculated center of mass shift of Moon be what it is?" Once one knows the answer to that question, then he or she will be able to simply deduce the crustal thickness difference.

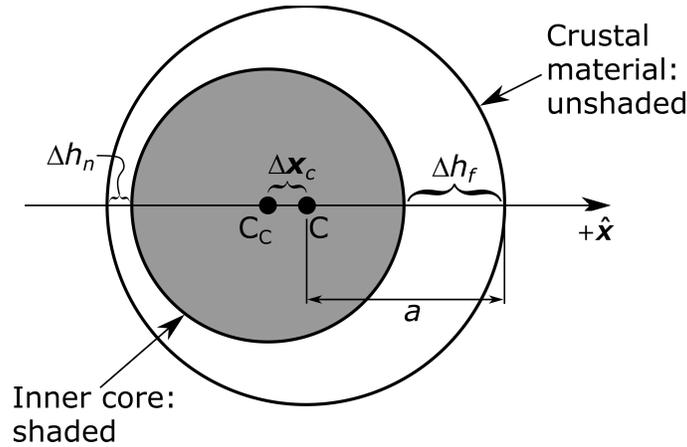

**Figure 4.** Shows two moon layers of different densities that produce the center of mass shift. The difference in crustal thickness is exaggerated to show the center of mass shift.

As in Figure 4 above, consider the Moon as being made up of two spheres with offset centers:

$C = $ The center of the Moon as a whole and Origin of the coordinate system, and

$C_C = $ The center and center of mass of the inner core

The overall sphere is made up of un-shaded crustal material on top of an assumed spherical inner core material. For the author's calculation, he assumes a constant average density for this sphere, whose value he will deduce.

Let the center of the inner core, $C_C$, be located a distance $\Delta x_c$ away from the center of the large sphere. The observed center of mass of the overall sphere consisting of both the inner core and the crustal material is located at $X_{CM}$. $X_{CM}$ is not shown in Figure 4.

*3.4. Solution*

Assume an imaginary crustal sphere all made out of crust. It consists of the crust on the outside plus crust replacing the inner core material as well. First, one may note that the center of mass of this imaginary crustal sphere occurs at the center of the overall sphere, at C. Next, he or she may subtract out the inner crust sphere whose shape matches the core's and add back in the spherical inner core. Both the imaginary crustal sphere that matches the shape of the inner core and the inner core sphere itself have center of mass at $\Delta x_C$.

How does $X_{CM}$ of the two-layer shape depend on $\Delta x_C$?

Let $V_1 = $ the volume of the smaller, inner core sphere with density $\rho_1$

Let $V = V_1 + V_2 = $ the total volume of the large sphere consisting of core volume $V_1$ and outer crust volume $V_2$. Then the crustal volume is:

$$V_2 = V - V_1 \tag{34}$$

The density of the crustal sphere is $\rho_2$

Using the definition of the center of mass:

$$\rho_2(V_1 + V_2)(0) - \rho_2 V_1(\Delta x_C) + \overline{\rho_1} V_1(\Delta x_C) = (\overline{\rho_1} V_1 + \rho_2 V_2)(X_{CM}) \tag{35}$$

Here the lunar mass $M_M$ is:

$$M_M = (\overline{\rho_1} V_1 + \rho_2 V_2) = \bar{\rho} V \tag{36}$$

One now solves for $\Delta x_C$:

$$\Delta x_C = \frac{(\overline{\rho_1} V_1 + \rho_2 V_2)}{(\overline{\rho_1} V_1 - \rho_2 V_1)} (X_{CM}) \tag{37}$$

Substituting $V_2 = V - V_1$ from equation (34) to eliminate the crustal volume and rewriting yields:

$$\Delta x_C = \frac{[(\overline{\rho_1} - \rho_2)V_1 + \rho_2 V]}{(\overline{\rho_1} - \rho_2)V_1}(X_{CM}) \tag{38}$$

Substituting (34) into (36) to eliminate $V_2$ and solving the resulting equation for the average inner core density, one has:

$$\overline{\rho_1} = \rho_2 + \frac{(\overline{\rho} - \rho_2)V}{V_1} \tag{39}$$

Next, substitute $\overline{\rho_1}$ from (39) into equation (38) to eliminate $\overline{\rho_1}$. One arrives at the following simple equation:

$$\Delta x_c = \left[\frac{\overline{\rho}}{(\overline{\rho} - \rho_2)}\right] X_{CM} \tag{40}$$

The author will now find an expression that relates $\Delta x_C$ to the crustal thickness difference between the far and near sides of the Moon.

Let $a$ = the Moon radius
Let $\Delta h_F$ = the crustal thickness on the far side
Let $\Delta h_N$ = the crustal thickness on the near side

The coordinates of the near and far sides of the inner core relative to the center C, which is also the Origin O, are:

$$-(a - \Delta h_N) \text{ and } (a - \Delta h_F)$$

This is because a coordinate is positive and equal to its distance from O if the coordinate is to the right of O and is equal to the opposite of its distance from O if the coordinate lies to the left of the origin.

Then $\Delta x_C$ can be related to the overall lunar radius, $a$, and the differences in crustal thickness, $\Delta h_N$ and $\Delta h_F$ as follows:

$\Delta x_C$ = The average of the near and far side coordinates

$$= \frac{-(a - \Delta h_N) + (a - \Delta h_F)}{2} \tag{41}$$

$$= \frac{\Delta h_N - \Delta h_F}{2} \tag{42}$$

Substituting $\Delta x_c$ from equation (42) into (40) and solving for $\Delta h_F - \Delta h_N$ yields:

$$\Delta h_F - \Delta h_N = -2 X_{CM} \left[\frac{\overline{\rho}}{(\overline{\rho} - \rho_2)}\right] \tag{43}$$

One may now plug the following density values into equation (43):

$$\rho_2 = 2691 \frac{kg}{m^3} \text{ [5] and}$$

$$\overline{\rho} = 3344 \frac{kg}{m^3} \text{ [9](229)}$$

In addition, one may plug in $X_{CM} = -1.57$ km, using equation (30), to arrive at:

$$\Delta h_F - \Delta h_N = 16.1 \text{ km} \tag{44}$$

To two significant figures:

$$\Delta h_F - \Delta h_N \approx 16 \text{ km} \tag{45}$$

The accepted average value for the difference in crustal thickness between the far and near sides of the Moon is 15 km [9] (p. 234). The percent error of the author is 6.3%, if two digits are used in the author's result. This is an excellent result considering the complex nature of the problem. The author understands that there are sources of error involved in this calculation that are hard for him to estimate, that when explored in full detail, may introduce a larger percent error, but he believes that this calculation is encouraging because it shows that his physical hypothesis that Archimedes' Law is the cause of the lunar crustal thickness difference and the lunar center of mass shift from center is likely to be true. That is, if the author's hypothesis and assumptions are true, then there must be a calculated crustal thickness difference. So the fact that this is indeed true and the value is approximately what was measured, makes his theory more likely. As far as the author can tell, the above calculation of crustal thickness difference between the far and near sides of the moon has never been done before based on a simple physical and mathematical model incorporating the same ideas as the author.

**4. Sources of Error**

1. The Moon and its inner core are not perfect spheres. If the inner core is not a sphere, then the center of mass of the inner core need not be located at the center of the inner core. The author's derivation would then have some error. The author assumes the inner core is approximately spherical for ease of calculation and because he believes it will not be very different from a sphere because the outside crust and differences in its thickness are so small relative to the core radius.

2. The Moon is not exactly in a circular orbit.

3. The expression for effective gravity is an approximation. The distance to the earth center from any point on the Moon is not quite equal to the distance represented by the component of displacement in the earth-moon center of mass line.

4. The center of mass of the Moon is not quite on the earth-moon line. It has been observed to lie at a tilting angle of 23 degrees off of it [7]. The deduced value in the earth moon line would then be the projected component of the center of mass in this direction. This deduced value would be smaller by the cosine of 23 degrees, which would make the actual predicted value of the center of mass in the line of cant approximately 1.7 km to two significant figures, reducing the error in the center of mass shift from 11% to only 5.6%.

5. The earth center is not exactly coincident with the earth-moon barycenter as the author assumed.

6. The author's calculation of lunar far-near crustal thickness difference is the value in the earth-moon line. It is not the average value. Inspection of Figure 4 shows that the author's value may be somewhat higher than the accepted average value.

7. The author uses his own calculated value for the center of mass shift to deduce the crustal thickness difference. If the accepted value were chosen then his calculated value of crustal thickness difference would be a higher value.

8. A possible complication not considered in the author's model is the difference in crustal rock. Both anorthosite highland rock and mare basalt rock are present in the lunar crust. The author takes the quoted average crustal density for his calculation [5].

**5. Conclusions and Implications**

The author has for the first time related the center of mass shift in the earth-moon line and the crustal thickness difference between the far and near sides in a simple physics and math model that gives results to within 5-10% of the accepted values. It also predicts a quantitative value for the decay of the lunar center of mass shift towards lunar center under lunar recession from earth. This result is explained nowhere else in the literature.

The author believes that there may be relations between the center of mass shift and the crustal thickness difference on other moons in the solar system or around other stars, given similar formation histories and parameters to our moon. In some circumstances, a moon's center of mass shift may increase or decrease with time, respectively, depending on whether the satellite moon approaches or recedes from its host. There may be opportunities to test the author's prediction by a lunar orbiter or an orbiting satellite around another moon around another planet. In the latter case, a moon could be chosen whose predicted value of center of mass shift were changing at a greater rate so as to make it more measurable by current technology. NASA missions like Cassini-Huygens and Galileo have explored the moons of Saturn and Jupiter, for example [19].

**Acknowledgements**

1. The author would like to thank Dr. Chris McKee of the UC Berkeley Astronomy and Physics Departments for helpful discussions on how to make his theory more quantitative. The idea of using averages to calculate the Moon's net effective gravity field and the idea of how to calculate the center of mass shift mathematically are due to Dr. McKee. Dr. Chris McKee may or may not advocate this theory.

2. The author would like to thank Kyoko Bischof for pointing out that he should read the article in *The American Scientist Magazine*, "The Two-Faced Moon." Without her, this paper would never have been written.

3. The author would like to thank the UC Berkeley librarians and staff at the Physics-Astronomy Library for helping him find some references and advising him on citation guidelines for online sources.

4. Technical illustrations drafted by David M. Akawie based on the author's sketches. Some editorial work by Akawie, as well.

**Funding:** The author received no sources of external funding for this research.

**Conflicts of Interest:** The author declares no conflicts of interest.